\documentclass[]{jpconf}
\usepackage{graphicx}
\begin{document}
\title{Measuring the muon's anomalous magnetic moment to 0.14 ppm}

\author{Frederick Gray, for the New ($g$-2) Collaboration\footnote{The New ($g$-2) Collaboration includes
 R.M.~Carey, K.R.~Lynch, J.P.~Miller, B.L.~Roberts (Boston University); 
 W.M.~Morse, Y.K.~Semertzidis (Brookhaven National Laboratory); 
 V.P.~Druzhinin, B.I.~Khazin, I.A.~Koop, I.~Logashenko, 
   S.I.~Redin, Y.M.~Shatunov, E.P.~Solodov (Budker Institute of Nuclear Physics);
 Y.~Orlov, R.M.~Talman (Cornell University);
 B.~Casey, B.~Drendel, K.~Genser, J.~Johnstone, A.~Jung, 
   D.~Harding, A.~Klebaner, A.~Leveling, J-F.~Ostiguy, N.V.~Mokhov, J.P.~Morgan, V.~Nagaslaev, 
   D.~Neuffer, A.~Para, C.C.~Polly, M.~Popovic, M.~Rominsky, A.~Soha, P.~Spentzouris, 
   S.I.~Striganov, M.J.~Syphers, G.~Velev, S.~Werkema (Fermi National Accelerator Laboratory);
 F.~Happacher , G.~Venanzoni, M.~Martini (Laboratori Nazionali di Frascati); 
 D.~Moricciani (Universit{\`a} di Roma ``Tor Vergata'');
 J.D.~Crnkovic, P.T.~Debevec, M.~Grosse-Perdekamp, 
   D.W.~Hertzog, P.~Kammel, N.~Schroeder, P.~Winter (University of Illinois at Urbana-Champaign);
 K.L.~Giovanetti (James Madison University);
 K.~Jungmann, C.J.G.~Onderwater (Kernfysisch Versneller Instituut);
 N.~Saito (KEK);
 C.~Crawford, R.~Fatemi, T.P.~Gorringe, W.~Korsch, B.~Plaster, 
   V.~Tishchenko (University of Kentucky);
 D.~Kawall (University of Massachussetts, Amherst);
 T.~Chupp, R.~Raymond, B.~Roe (University of Michigan);
 C.~Ankenbrandt, M.A.~Cummings, R.P.~Johnson, C.~Yoshikawa (Muons Inc.)
 A.~de Gouv{\^{e}}a (Northwestern University);
 T.~Itahashi, Y.~Kuno (Osaka University);
 G.D.~Alkhazov, V.L.~Golovtsov, P.V.~Neustroev, 
   L.N.~Uvarov, A.A.~Vasilyev, A.A.~Vorobyov, M.B.~Zhalov (Petersburg Nuclear Physics Institute);
 F.~Gray (Regis University);
 D.~St{\"{o}}ckinger (Technical University of Dresden);
 S.~Bae{\ss}ler, M.~Bychkov, E.~Frle{\u{z}}, and D.~Po{\u{c}}ani{\'{c}} (University of Virginia).
}}

\address{Department of Physics and Computational Science, Regis University, Denver, CO 80221, USA}

\ead{fgray@regis.edu}

\begin{abstract}
The anomalous magnetic moment ($g$-2) of the muon was measured with a precision of 0.54 ppm in Experiment 821 at Brookhaven National Laboratory.  A difference of 3.2 standard deviations between this experimental value and the prediction of the Standard Model has persisted since 2004; in spite of considerable experimental and theoretical effort, there is no consistent explanation for this difference.  This comparison hints at physics beyond the Standard Model, but it also imposes strong constraints on those possibilities, which include supersymmetry and extra dimensions.  The collaboration is preparing to relocate the experiment to Fermilab to continue towards a proposed precision of 0.14 ppm.  This will require 20 times more recorded decays than in the previous measurement, with corresponding improvements in the systematic uncertainties.  We describe the theoretical developments and the experimental upgrades that provide a compelling motivation for the new measurement.
\end{abstract}

\section{Introduction}

\begin{figure}
\includegraphics[width=\textwidth]{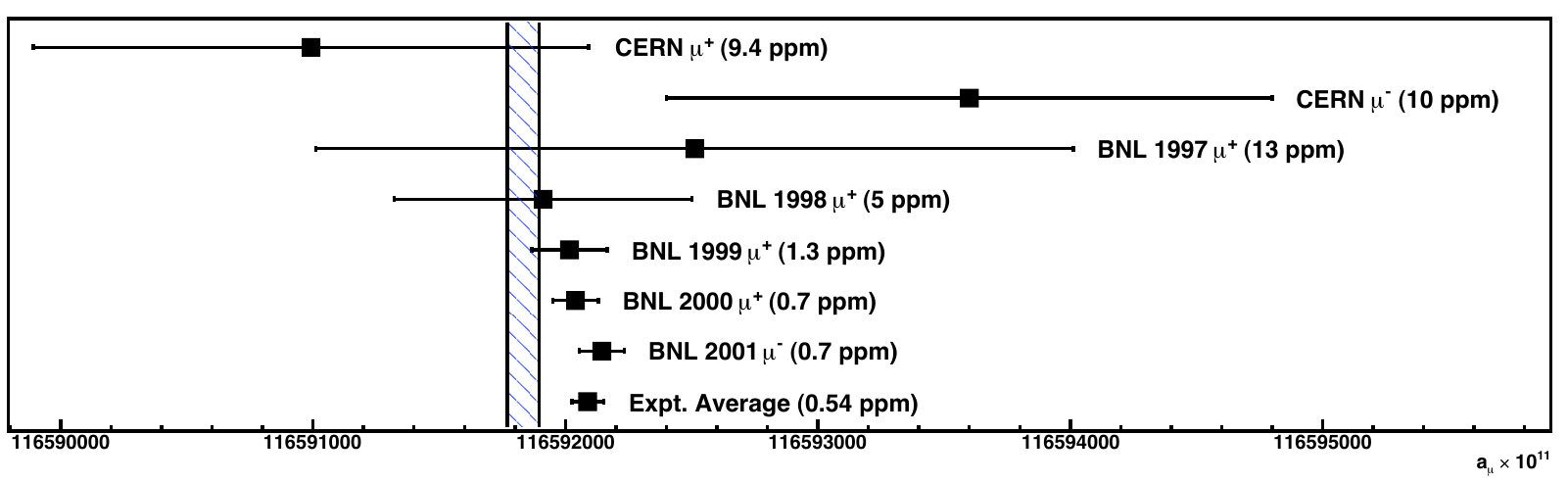}
\caption{History of the precision of the measurements of $a_\mu$ in the CERN III experiment~\cite{Bailey:1978mn} and
in E821, compared with the Standard Model prediction~\cite{Davier:2009zi}.}
\label{history}
\end{figure}

The principle of the muon ($g$-2) experiment is to store a polarized beam of muons in a magnetic field and
observe the precession of the muon spin direction with respect to the muon momentum.   This precession occurs
at an angular frequency $\omega_a = \frac{e}{m} a_{\mu} B$, where $e$ is the muon charge, $m$ is its
mass, $a_{\mu}$ is the anomalous magnetic moment, and $B$ is the applied magnetic field.
To determine the anomalous magnetic moment 
with great precision,
$\omega_a$ and $B$ must be measured very precisely.  

Parity violation in muon decay ($\mu^\pm \rightarrow e^\pm + \nu + \bar\nu$) causes higher-energy
positrons to preferentially follow the muon spin direction in the muon rest frame for $\mu^+$ decay,
and higher-energy electrons to preferentially go opposite to the muon spin for $\mu^-$ decay.
In the laboratory frame, the electron energy is greatest when the muon spin points
forward, along the muon momentum direction.   The rate of high-energy electron events in calorimeters
placed adjacent to the muon storage ring is therefore modulated at the precession frequency $\omega_a$.  

The magnetic field is monitored using a nuclear magnetic resonance (NMR) technique,
observing the rate of spin precession of the free protons in a large collection of probes.
A ``trolley'' of NMR probes is driven around the muon storage region every few days.
Variations between trolley runs are monitored by fixed probes located just outside of it.
The $B$ field must be averaged over the spatial distribution of muons.  
Careful shimming makes the magnetic field as uniform as possible over the muon storage region
to reduce the dependence on the muon distribution.

The magnetic dipole moment of a particle is proportional 
The gyromagnetic ratio $g$ of a particle relates its magnetic dipole moment $\vec\mu$ to its spin $\vec S$
by $\vec\mu = g \left( \frac{e}{2m} \right) \vec{S}$.  .  
For a pointlike Dirac particle, $g = 2$.
The anomalous magnetic moment $a_\mu = \frac{1}{2} (g-2)$ describes substructure
and coupling to virtual fields; while the nearly pointlike electron and muon have $g \approx 2.002$,
the proton has $g \approx 5.586$, a reflection of its internal structure.
For the muon, it can be calculated theoretically in the context of the Standard Model with very high precision.  
A comparison of the experimental and theoretical values tests the Standard Model as a whole
and is potentially sensitive to extensions such as supersymmetry, which would generally contribute~\cite{Czarnecki:2001pv} 
\begin{displaymath}
a_{\mu(SUSY)} = {\rm{sgn}}(\mu) (130 \times 10^{-11}) \tan\beta \left( \frac{\rm 100~GeV}{\Lambda} \right)~.
\end{displaymath}
Therefore, precise measurements of a low-energy system, with energy scales from MeV to GeV, may show 
small effects from interesting high-energy processes at nearly the TeV scale.  

\section{Status of experiment and theory}

At Brookhaven National Laboratory (BNL), Experiment E821 at the Alternating Gradient Synchrotron (AGS) 
measured $a_\mu$ to a precision of 0.54 parts per million (ppm)~\cite{Bennett:2006fi}, 
$a_{\mu;expt} = (116\,592\,089 \pm 63) \times 10^{-11}$.
The Standard Model prediction~\cite{Davier:2009zi} has a precision of 0.42 ppm, 
$a_{\mu;SM} = (116\,591\,834 \pm 49) \times 10^{-11}$.
Consequently, a difference of $a_{\mu;expt} - a_{\mu;SM} = 255 \pm 80$ exists between the current experimental value and
theoretical prediction.  While this difference of 3.2 standard deviations is not yet a ``discovery,'' $a_\mu$ is the only
low-energy observable that 
cannot be fully explained by the Standard Model.  The probability that such a difference would appear by chance
is at the level of $10^{-3}$.
This difference has persisted at nearly the same level since the final result 
from E821 was published in 2004 in spite of significant improvements in the Standard Model calculation. 
The series of experimental measurements from each year's run appears in Figure~\ref{history} with a
comparison to the Standard Model prediction.

The Standard Model calculation is dominated by terms from quantum electrodynamics; 
this part has been carried out through four-loop processes, requiring the evaluation of more 
than 1000 diagrams, and giving $a_{\mu;QED} = (116\,584\,718.09 \pm 0.16) \times 10^{-11}$~\cite{Kinoshita:2005zr},
a precision of 0.001~ppm.  Electroweak processes add $154 \pm 2 \times 10^{-11}$~\cite{Czarnecki:2002nt},
a precision of 0.02~ppm.  These contributions are well-understood and do not contribute 
significantly to the theoretical uncertainty, which arises almost exclusively from hadronic 
vacuum polarization and hadronic light-by-light scattering, which contribute 
$a_{\mu;HVP} = (6\,857 \pm 41) \times 10^{-11}$~\cite{Davier:2009zi,Hagiwara:2006jt}
and $a_{\mu;HLBL} = (105 \pm 26) \times 10^{-11}$~\cite{Prades:2009tw}
respectively.

The theoretical contribution from hadronic vacuum polarization cannot be calculated analytically
with high precision, 
so this term must be estimated from experimental data.   It may be related through a
dispersion integral to the cross section for real hadron production in $e^+ e^-$ collisions,
normally expressed as a ratio $R(s)$ to the cross section for muon production. 
The dispersion integral weights the data by $1/s^2$, emphasizing the low-energy region 
up to about 1~GeV, where a two-pion final state is dominant.
At the time of the initial publication of the E821 results, the only precise measurements
of $R(s)$ at many energies in this range were from the CMD-2 experiment
at the VEPP-2 collider in Novosibirsk~\cite{Akhmetshin:2003zn}.
Subsequently, these measurements have been confirmed and improved
by further work with CMD-2~\cite{Aulchenko:2006na,Akhmetshin:2006wh,Akhmetshin:2006bx},
by its companion experiment SND~\cite{Achasov:2006vp},
and using  an initial-state radiation technique 
at BaBar~\cite{:2009fg} and KLOE~\cite{:2008en}.  Agreement at the level of two standard deviations is now seen 
throughout this critical energy region.  Meanwhile, even though isospin-breaking corrections
have been improved, substantial local discrepancies remain between this analysis and
the corresponding cross sections that are inferred from hadronic $\tau$ decay rates~\cite{Davier:2009zi},
raising further doubts about the reliability of the $\tau$ decay method.

The final result from E821 remained limited by statistics, as shown in Table~\ref{systematicTable},
so an improvement in precision would be possible simply with increased running time.  Unfortunately,
high-intensity proton operations are no longer feasible at BNL.  Consequently, 
a proposal~\cite{Carey:2009zzb} has been developed to relocate the E821 apparatus to Fermi National Accelerator 
Laboratory (Fermilab) to collect approximately 20 times more muon decays than are in the 
current data set, with corresponding improvements in systematic uncertainties
to reach a combined uncertainty of 0.14~ppm.

\section{A new experiment at Fermilab}

\begin{table}
\begin{tabular}{llll}
\hline
  & Run 2000 ($\mu^+$) & Run 2001 ($\mu^-$) & Fermilab target \\
\hline
Statistics & 0.62 & 0.67 & 0.10 \\
\hline
Systematics in $\omega_a$: & 0.31 & 0.21 & 0.07 \\
{- Overlapping pulses (pileup)} & 0.13 & 0.08 & 0.04 \\
{- Detector gain changes} & 0.13 & 0.08 & 0.04 \\
{- Muon losses} & 0.10 & 0.09 & 0.02 \\
{- Coherent betatron oscillations} & 0.21 & 0.07 & 0.04 \\
\hline
Systematics in $\omega_p$ ($B$ field) & 0.24 & 0.17 & 0.07 \\
\hline
\end{tabular}
\caption{Estimates of statistical and systematic uncertainties from E821 and the future Fermilab experiment;
all values are in parts per million (ppm).  ``Run 2000'' refers to the positive muon result published in~\cite{Bennett:2002jb} 
and ``Run 2001'' to the negative muon result in~\cite{Bennett:2004pv}.}
\label{systematicTable}
\end{table}

The Fermilab accelerator complex will be able to deliver a pulsed muon beam to the experiment with a higher total intensity, 
but subdivided into more bunches and with less pion contamination than the AGS.   While the NOvA experiment~\cite{Ayres:2004js} 
is running with the Booster in 15~Hz mode, eight out of twenty Booster batch cycles will be available for other uses, and 
six of these can be directed to the ($g$-2) experiment.
They will be extracted into the
Recycler, which will be used to subdivide each batch into four smaller bunches.  From the Recycler, these bunches will 
be extracted one at a time into the target area currently used to produce antiprotons, and the long beamline to and from 
the antiproton storage ring will be used as a pion decay channel.  

A new experimental hall will be constructed near the antiproton source.
The four coils of the superconducting magnet have a diameter of 14~m and therefore cannot be transported 
by road from BNL.  A helicopter crane will move them to a barge waiting in the Long Island Sound, which will transport them 
through the St.~Lawrence Seaway and the Great Lakes to the Calumet-Saganashkee Channel in Illinois.  
A helicopter crane will again move them to the new site at Fermilab.  

The efficiency for producing stored muons at Fermilab (the number of stored muons normalized to the number
of protons on target) is expected to be at least 6 times higher than it was at the AGS.
The pion yield per proton on target will be lower by a factor of 2.5 for the 8~GeV protons from the Fermilab Booster
than it had been with the 24~GeV proton beam from the AGS, but every other factor provides an increase.
A larger pion momentum range, $\pm$2\% rather than $\pm$0.5\%, will be accepted by the beamline.
The pion decay channel will have a length of approximately 900~m rather than 88~m.
All forward pion decays will be accepted, rather than only decays into a 3.8~mr solid angle.
The quadrupole magnets will be spaced at intervals of 3.25~m rather than 6.2~m, resulting in more effective
focusing and fewer losses in the transport line.  Finally, the superconducting inflector magnet~\cite{Yamamoto:2002bb}
that shields the beam from the main storage ring magnetic field as it is injected will be rebuilt with open ends;
in E821, the conductor was wound around both ends, causing significant multiple scattering.
The repetition rate of muon bunches will be 18~Hz, to be compared with 4.4~Hz at BNL,
and each bunch will contain $\sim10^{4}$ muons.  Consequently, it will be possible to collect the required
statistics of $2\times10^{11}$ accepted, analyzed decays in less than two years of running time.

The E821 analysis effort was complicated by the presence of a ``flash'' background from neutrons produced
through spallation by pion contamination in the beam.  These neutrons thermalized and were captured
in the acrylic lightguides of the electron calorimeters, resulting in a slow baseline shift in their signals.
This baseline shift complicated the analysis of several important systematic errors, leading to
estimates of the effects of overlapping pulses (pileup), calorimeter gain changes, and muon losses 
that were probably overly conservative.  Primarily because the decay channel will be much longer,
the number of pions in each fill is expected to be reduced by a factor of 20 in the Fermilab experiment, 
and the ``flash'' should be reduced by this same factor.

In addition, in order to deal with the increased muon rate, new electron calorimeters are being designed with greater 
segmentation.  The previous lead/scintillating fiber calorimeters~\cite{Sedykh:2000ex} were divided into four blocks whose
signals were summed before digitization.  The new tungsten/scintillating fiber calorimeters~\cite{McNabb:2009dz}
will have at least 35-fold segmentation, and each segment will be digitized separately by new
waveform digitizers and a new data acquisition system.  Consequently, it will be possible to 
separate events that overlap in time but not in space.
Prototypes of the new calorimeter have been constructed and tested at the
Meson Test Beam at Fermilab, and the GEANT4 simulation of the device was validated by a measurement
of the energy resolution.

Coherent betatron oscillations of the muon beam modulated the rate, asymmetry, and phase of the 
distribution of detected electrons at a frequency only slightly different than $2 \omega_a$,
leading to an important systematic error.  Several options are being studied to suppress
these oscillations in the new experiment, including a pulsed octupole field or an oscillating
dipole field that would be applied just after the beam is injected.

The precision of the magnetic field measurement will be improved by continued refinement
of the careful shimming and calibration techniques demonstrated in E821.  
It will be necessary to measure the field perturbations caused by eddy currents from the muon injection 
kicker and by the fringe field of the new open-ended inflector.
The number of fixed NMR probes will be increased, they will be moved to locations where they are more effective, 
and trolley maps of the field will be collected more frequently.  

\section{Conclusion and outlook}

The BNL E821 experiment has measured the muon's anomalous magnetic moment to a precision
of 0.54~ppm, comparable to that of the Standard Model theory value, which has a precision of
0.42~ppm.  The 3.2 standard deviation difference between these values may provide a hint of 
new physics; while it does not constitute a discovery, it provides a strong motivation 
for an improved measurement.
A clear opportunity exists to improve the experimental precision to 0.14~ppm by relocating 
to Fermilab.  There is a finite window of time to do this, while the collaboration and the magnet remain
mostly intact.

There is a clear path to improve the precision of the Standard Model prediction to $\sim$0.25~ppm
over the next decade using existing and anticipated data on hadron production 
in $e^+ e^-$ collisions.  When this prediction is subtracted from result of the upgraded ($g$-2) experiment,
the precision of the difference will be 0.3~ppm, to be compared with 0.7~ppm today.
Consequently, the combination of our new measurement with this improved Standard Model value 
will provide a unique opportunity to constrain the interpretation of any new physics discovered 
at the LHC.

\section*{References}

\bibliographystyle{iopart-num}

\bibliography{fgray_inpc}

\end{document}